\documentclass{aa}

\usepackage[utf8]{inputenc}
\usepackage[T1]{fontenc}

\usepackage{graphicx}
\usepackage{dcolumn}
\usepackage{bm}
\usepackage{newtxtext,newtxmath}
\usepackage{amssymb}
\usepackage{amsmath}
\usepackage{amsfonts, dsfont}
\usepackage{graphicx}
\usepackage{hyperref}
\hypersetup{colorlinks=true, citecolor=teal, urlcolor=teal, linkcolor=teal}

\usepackage{accents}
\usepackage{color}
\usepackage[dvipsnames]{xcolor} 
\usepackage{lineno}
\usepackage{xspace}
\usepackage[normalem]{ulem} 
\usepackage{float}
\usepackage{orcidlink}
\usepackage{siunitx}
\newcommand{\gwb}{\mathrm{GWB}}
\newcommand{\intl}{\int\limits}

\newcommand{\toa}{\delta t}
\newcommand{\mtx}[1]{\mathbf{#1}}

\newcommand{\mpifr}{Max-Planck-Institut f{\"u}r Radioastronomie, Auf dem H{\"u}gel 69, 53121 Bonn, Germany}
\newcommand{\SPA}{School of Physics and Astronomy, Monash University, Clayton, VIC 3800, Australia}
\newcommand{\OzGravMonash}{OzGrav: The ARC Centre of Excellence for Gravitational Wave Discovery, Clayton, VIC 3800, Australia}

\newcommand{\jbca}{Jodrell Bank Centre for Astrophysics, University of Manchester, Department of Physics and Astronomy, Alan-Turing Building, Oxford Street, Manchester M13 9PL, UK}
\newcommand{\aei}{Max-Planck-Institut für Gravitationsphysik (Albert-Einstein-Institut), Callinstraße 38, 30167 Hannover, Germany}
\newcommand{\potsdam}{Leibniz Universität Hannover, 30167 Hannover, Germany}

\begin{document}

\title{Not just a phase: detecting nanohertz gravitational waves from phase alone}

\author{Kathrin Grunthal \orcidlink{0009-0000-2963-6641}\inst{1}\thanks{kgrunthal@mpifr-bonn.mpg.de}  \and
        Eric Thrane \orcidlink{0000-0002-4418-3895}\inst{2,3}  \and
        Rutger van Haasteren \orcidlink{0000-0002-6428-2620}\inst{4,5} \and
        David J. Champion \orcidlink{0000-0003-1361-7723}\inst{1} \and
        Michael Kramer \inst{1,6}
        }

\institute{
    \mpifr \and 
    \SPA \and
    \OzGravMonash \and
    \aei \and
    \potsdam \and
    \jbca
}

\date{\today}

\abstract
    {Unambiguously detecting a gravitational-wave background in pulsar-timing data requires a robust detection statistic. The pulsar timing community currently relies heavily on the cross-correlation-based ``optimal statistic.'' Its design was motivated to avoid false-positive detections from unmodeled pulsar red noise, which, it was thought, can be mitigated by searching for signals that are coherent across multiple pulsars. However, recent work by van~Haasteren (2025) suggests that the optimal statistic is not, in fact, optimal. Moreover, the optimal statistic is still susceptible to false positives from misspecified noise. To guard against this, analysts employ quasi-resampling methods such as phase shifts, in which complex phases are randomly assigned to each pulsar strain measurement, thereby destroying any gravitational-wave-induced correlations. The shifted data is then used to approximate the null distribution for the optimal statistic.}
    {Inspired by the concept of phase shifts, we introduce a formalism in which the gravitational-wave background is measured entirely from phase measurements. We argue that this approach is robust against false positive detections without the need for ad hoc quasi-resampling methods.}
    {We demonstrate the feasibility of this formalism by measuring the amplitude and spectral index of a simulated gravitational-wave background signal, injected to a simplified mock representation of a pulsar timing array data set. For this example, we use simulation-based-inference and model the phase-based likelihood using a neural circular spline flow. 
    By comparing our results to ones obtained with the optimal statistic, we show how much resolving power comes from amplitude versus phase.}
    {We show that it is possible to adequately recover the spectral parameters of the gravitational wave background signal using the phase-based likelihood.  By comparing our results to ones obtained with the optimal statistic, we show how much resolving power comes from amplitude versus phase.}
    {}

\keywords{gravitational waves - methods: data analysis - methods: statistical - pulsars: general - stars: black holes}

\maketitle

\section{Introduction}
\label{sec:intro}

Pulsar timing arrays (PTAs) seek to detect nanohertz gravitational waves (GWs) by monitoring the highly regular pulsar arrival times of Galactic millisecond pulsars using radio telescopes. Such low-frequency GWs are expected to either originate from supermassive black hole binaries, formed in galaxy mergers \citep{Kormendy_1995,Begelman_1980}, or more exotic processes such as phase transitions, collapsing cosmic strings etc.\ that might have happened during the early phases of the Universe \citep[e.g.,][]{Hindmarsh_1995,Hindmarsh_2014,Caprini_Figueroa_2018,Domenech_2021}. The most anticipated signal to be detected in the frequency range of PTAs is a stochastic gravitational wave background (GWB), formed by the incoherent superposition of a large number of individual nanohertz GWs \citep{Rajagopal_1995}. When the radio pulses are influenced by such a GW signal, the resulting fluctuations of the pulse arrival times are correlated between pairs of pulsars according to the Hellings-Downs (HD) correlation function \citep{HellingsDowns_1983}. 

Multiple PTA collaborations \citep{EPTA_DR2_GWB,PPTA_DR3_GWB,NANOGrav15yr_GWB,MPTA_4.5yr_GWB,CPTA_DR1} have reported evidence for such an Hellings-Downs-correlated signal in their data set. However, as the reported false alarm probabilities (FAPs) for the presence of this signal range between $2$-$4\sigma$, these results are for the time being short of the canonical $5\sigma$ threshold \citep{Allen_2023_detectionthreshold}. 
In order to establish the significance of these signals, the PTA community relies on the so-called  ``pair-correlation optimal statistic'' (or ``optimal statistic'') \citep{Demorest_2013,Chamberlin_2015}, which is an average of pairwise pulsar correlations, weighted by a filter which accounts for the noise properties of the data. We provide an overview of the optimal statistic in Appendix~\ref{app:pcos}.

This resorting to pair-wise cross correlations (instead of incorporating all available information from pulsar-timing measurements) lies in the community's concern, that pulsar noise models are likely misspecified. If the PTA community trusted the pulsar noise models, sufficiently large excess red noise could be considered a smoking-gun signal of gravitational waves. However, it is seen as insufficient evidence to claim a gravitational-wave detection. 

As a result, the community insists that the detection statistic is a weighted sum of cross-correlation, which, in the weak-signal limit \cite{vanHaasteren_2025p}, results in the optimal statistic as described in Appendix~\ref{app:pcos}. In other words, the community adopted the optimal statistic---which intentionally throws away some information---in order to guard against false positives from noise misspecification.

Taking a closer look however, it is, by some measures, \textit{sub-optimal}. Not only, because it does not account for all information from the observations, but also, because it is still sensitive to noise misspecification, in spite of this compromise.
The ``optimal signal-to-noise ratio'' (S/N; see Appendix~\ref{app:pcos}) depends on the choice of pulsar noise models, which are fit with phenomenological models.
The S/N is not taken at face value because misspecified noise can still lead to outliers from the null distribution.
As a result, PTA analyses use quasi-resampling techniques such as sky scrambling \citep{CornishSampson_2016} and phase shifting \citep{Taylor_2017} in order to convert the optimal S/N into a frequentist $p$-value \citep{DiMarco_2023,vanHaasteren_2025p}. 
While significances estimates from quasi-resampling are probably reliable for the purposes of detection \cite{DiMarco_2024}, this workflow leads to a disconnect where detection and inference of astrophysical parameters are not carried out within a single unified framework.

In this paper we ask the question: can we construct a formalism with the following properties?
\begin{enumerate}
    \item It is safe from false positives even when pulsar noise is misspecified.
    \item It does not rely on quasi-resampling to assess significance.
    \item It provides the most statistically powerful test possible with the information that we trust.
\end{enumerate}
For inspiration, we turn to phase scrambling, in which complex phases are randomly assigned to each pulsar strain measurement \cite{Taylor_2017}.
Phase scrambles spoil the coherence of a gravitational-wave signal, while preserving the amplitude, allowing pulsar timers to estimate the null distribution of the optimal statistic in a way that is less affected by misspecified noise.
Phase scrambles are based on the premise that the amplitude information is completely untrustworthy while the phase information is entirely reliable.

We follow this logic to its natural conclusion and construct a likelihood that depends only the phase of pulsar-timing strain measurements.
We show that one can measure the amplitude of the gravitational-wave background by looking for correlations between phases in pulsar strain measurements.
We show that one can measure the spectral index of the background by measuring how these correlations change with frequency.
We suggest that the resulting framework meets the three criteria listed above, avoiding false positives without the need for quasi-resampling methods.
By comparing our results to those obtained with the optimal statistic, we assess the extent to which the optimal statistic uses the less trustworthy amplitude information.

The remainder of this paper is organized as follows. In Sec.~\ref{sec:fourier_timing_residuals} we develop a formalism for the PTA phase likelihood. We then implement this approach using simulation-based inference with Monte Carlo data, as detailed in Sec.~\ref{sec:demonstration}. Finally, we discuss and conclude our results in Sec.~\ref{sec:summary_outlook}.

\section{Pulsar timing array formalisms}

A PTA data set consists of $N_\mathrm{psr}$ time series of $N_\mathrm{ToA}$ timing residuals. Under a Fourier transform, they become $N_\mathrm{psr}$ vectors of complex values $\vec{z}$, with length $N_\mathrm{bin}$, where $N_\mathrm{bin}$ is the number of frequencies used for the discrete Fourier transform. Under the assumption that any deterministic signal has been correctly modelled by the timing model, the residuals contain only contributions from stochastic signals. This includes the GWB signal, $\vec{s}_\gwb$, but also any noise processes, $\vec{n}$, such as instrumental white noise, or the plethora of uncorrelated red noises encountered in pulsar timing, such as timing noise, dispersion measure noise or scintillation noise \citep{Lentati_2013}.

\subsection{Fourier domain noise and signal models}
\label{sec:fourier_timing_residuals}

The basic assumption of all PTA signal modelling is, that all random signals can be modelled as Gaussian processes with zero mean and some noise-dependent covariance \citep{Lentati_2013}. It is convenient to express the frequency-domain realisations of these signals in terms of their real and imaginary parts, $u_X$ and $v_x$, respectively, as
\begin{align}
    \vec{z} &= \vec{s}_\gwb &&+ \vec{n} \\
    &\equiv \vec{u}_\gwb + i\, \vec{v}_\gwb &&+ \vec{u}_\mathrm{n} + i\, \vec{v}_\mathrm{n}.
\end{align}
The noise contribution of the $a^\mathrm{th}$ pulsar follows $\vec{u}_{\mathrm{n},a}, \vec{v}_{\mathrm{n},a}\sim\mathcal{N}(0,\sigma_a)$, with
\begin{equation}
    \sigma_a = \sqrt{\mathcal{I}^2\sigma^2_\mathrm{ToA} + \mathcal{Q}^2 } \,\frac{2N_\mathrm{bin}+1}{\sqrt{2N_\mathrm{bin}}},
\end{equation}
where $\mathcal{I}$ and $\mathcal{Q}$ denote the white noise parameters EFAC and EQUAD, respectively, and $\sigma_\mathrm{ToA}$ is the ToA uncertainty.\footnote{In our demonstration we assume that EFAC and EQUAD are known precisely, but in principle, one should marginalize over uncertainty in these parameters. We return to this approximation in Section~\ref{sec:summary_outlook}.} The factor $\frac{2N_\mathrm{bin}+1}{\sqrt{2N_\mathrm{bin}}}$ originates from the conversion of the time-domain definition of white noise to the frequency domain.
The GWB contribution is modelled jointly across all pulsars, with $\vec{u}_\gwb, \vec{v}_\gwb\sim\mathcal{N}(0,\mathbf{S})$, and 
covariances of the GWB and noise in Fourier domain are given as 
\begin{equation}\label{eq:residuals_ccmatrix}
    S_{ab} = \Gamma_{ab}\, \frac{1}{2} S(f)\, \Delta f,
\end{equation}
with the Hellings-Downs coefficient \citep{HellingsDowns_1983}, $\Gamma_{ab}$, the frequency difference between two Fourier frequencies, $\Delta f$, and the frequency-dependent GWB spectrum,
\begin{equation}
    S_\gwb(f) = \frac{A_\gwb^2}{12\pi^2}\left(\frac{f}{f_\mathrm{yr}}\right)^{-\gamma_\gwb} f_\mathrm{yr}^3,
\end{equation}
which is modelled as a powerlaw governed by its amplitude, $A_\gwb$, and spectral index, $\gamma_\gwb$.
The full concatenated complex residuals are denoted $\vec{\mathcal{Z}} \equiv (\vec{z}_1, \ldots, \vec{z}_{N_\mathrm{psr}})^T$.\newline \newline

At this point, we would like to stress, that this work concerns two different amplitudes with two different roles. Regarding the description of the data set using complex numbers written as $z=\mathcal{A} e^{i\phi}$, we mention the amplitudes, $\mathcal{A}$, of these complex numbers (for example, above, in describing the phase scrambling). With respect to the GWB signal, we describe its powerlaw power spectral density in terms of its amplitude, $A_\gwb$, and spectral index, $\gamma_\gwb$. Whenever the GWB amplitude is meant, it is clearly denoted by the subscript, and we aim to distinct the complex amplitude by resorting to a different font.

\subsection{Timing-residual phase likelihood function}
\label{ssec:phase_likelihood_function}
Assuming that the aforementioned noise models are appropriately representing the noise budget of the PTA data set, the likelihood for the residuals reads
\begin{equation}\label{eq:residual_pdf}
    p(\vec{\mathcal{Z}}|A_\gwb, \gamma_\gwb) = \frac{1}{\pi^{2N} \sqrt{\det{\Sigma}}} \exp\left[-\vec{\mathcal{Z}}^\dagger \mathbf{\Sigma}^{-1} \vec{\mathcal{Z}} \right],
\end{equation}
where the covariance matrix $\mathbf{\Sigma}$ is constructed from the noise models, and $N=N_\mathrm{psr}\cdot N_\mathrm{bin}$
Assuming that the data sets only contains a powerlaw GWB signal and fixed white noise, it is block diagonal and reads
\begin{equation}\label{eq:sigma}
    \Sigma_{ab, kk'} = \delta_{kk'}\left(\Gamma_{ab}\, S_\gwb(f_k)\, \Delta f  \,+ \delta_{ab}\,\sigma_a^2\right)
\end{equation}

If we rewrite the complex-valued residuals in terms of amplitudes and phase,
\begin{equation}\label{eq:phase}
z = \mathcal{A} e^{i\phi} ,
\end{equation}
the likelihood becomes a joint distribution of amplitudes and phase given the parameters $ A_\gwb, \gamma_\gwb$:
\begin{align}\label{eq:llh_dep_Aphi}
    {\cal L}(\vec{\mathcal{A}}, \vec{\phi} | A_\gwb, \gamma_\gwb) .
\end{align}
The likelihood of $(A_\gwb, \gamma_\gwb)$ that solely depends on the phases, $\{\vec{\phi}\}$, is then obtained by marginalising Eq.~\ref{eq:llh_dep_Aphi} over the amplitudes, $\{\vec{\mathcal{A}}\}$.
Defining
\begin{equation}\label{eq:M}
    \mathbf{M}(\vec{\phi}) \equiv e^{-i \vec{\phi}}\,\mathbf{\Sigma}^{-1} e^{i \vec{\phi}} ,
\end{equation}
the resulting phase-dependent likelihood reads
\begin{multline}\label{eq:phase_llh}
    \mathcal{L}(A_\gwb, \gamma_\gwb|\vec{\phi}) =\\ \frac{1}{\pi^{2N} \sqrt{\det{\Sigma}}}\intl_0^\infty\!\rm{d}\vec{\mathcal{A}} \,\left(\prod_i\mathcal{A}_i\right)\exp\left[-\vec{\mathcal{A}}^T\; \mathbf{M}(\vec{\phi})\; \vec{\mathcal{A}} \right],
\end{multline}
where the data enters via $\mathbf{M}(\vec{\phi})$.

In the presence of a correlated signal, $\mathbf{\Sigma}$ contains non-zero entries on the off-diagonal, which leads to correlations between phases. In turn, $\mathbf{M}(\vec{\phi})$ is similarly not diagonal, but contains non-zero off-diagonal elements, that depend on the differences between different phases, $\Delta\vec{\phi}$.
The marginal distributions for each phase are still uniform, but the joint distribution of the phases is not uniform.\footnote{Joint distributions with uniform marginal distributions that allow for covariance between parameters are known as copulas. Thus, the distribution of phases forms a high-dimensional copula. See \cite{Adamcewicz_2022} for a discussion of copulas in the context of gravitational-wave astronomy}

\begin{figure}
    \centering
    \includegraphics[width=\linewidth]{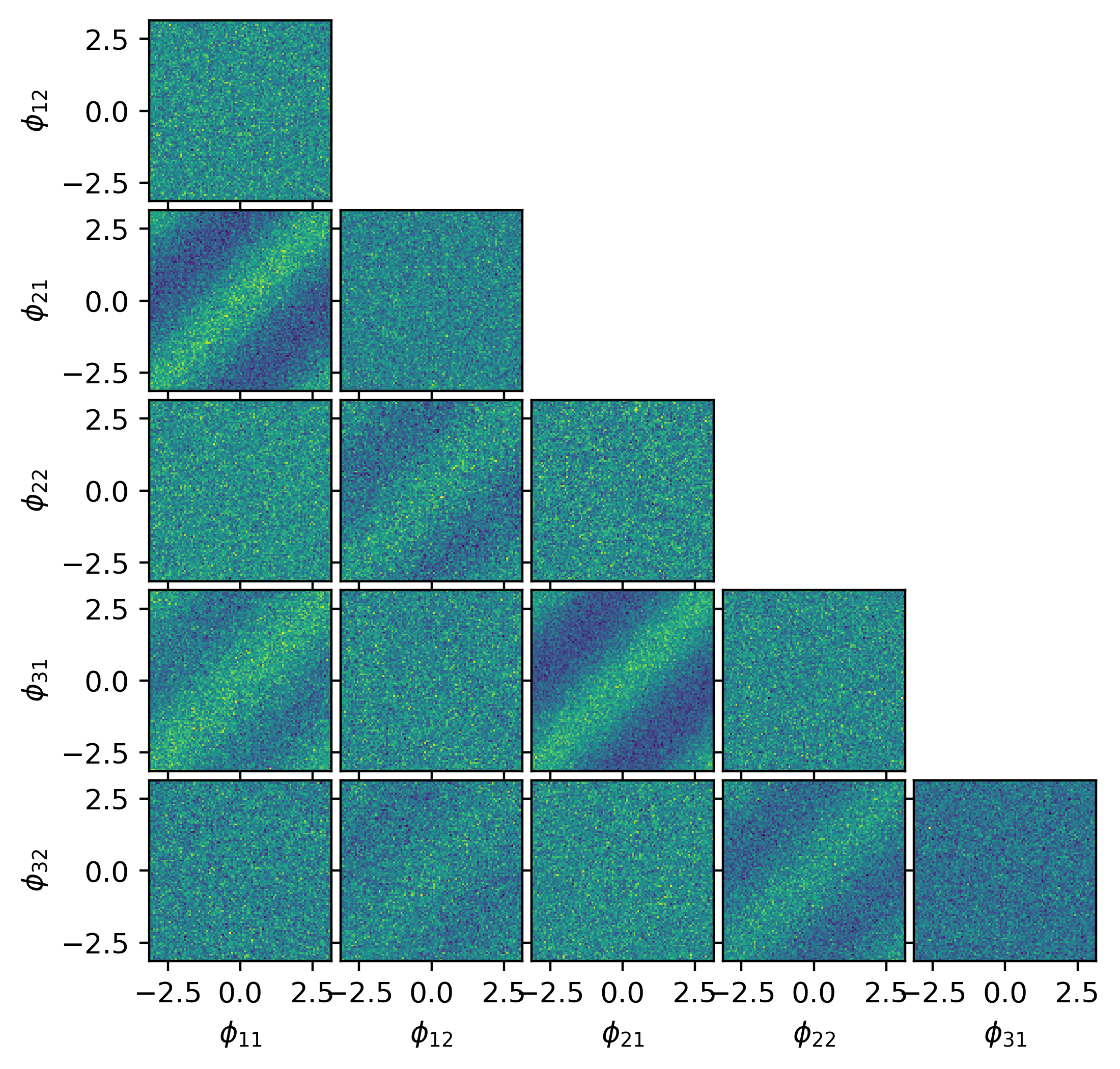}
    \caption{The phase likelihood distribution of for three pulsars and two frequency bins. The data contain white noise and an isotropic gravitational-wave background.}
    \label{fig:phasedistribution_example_2p_2b}
\end{figure}

For two pulsars and two frequency bins, with statistically independent noise, the empirical likelihood distribution is shown in Fig.~\ref{fig:phasedistribution_example_2p_2b}. Focusing on panels corresponding to two phases in the same frequency bin, we observe two stripes at a \SI{45}{\degree} angle. The visibility of this pattern decreases with increasing frequency because the correlations stand out less above the noise.
They are also more or less pronounced depending on the angular separation of the pulsars.

Yet, for a PTA, $\mathbf{\Sigma}$ is neither diagonal nor block-diagonal. Therefore the integral in Eq.~\ref{eq:phase_llh} only has an exact analytical solution for $N\leq2$ pulsars. While the analysis of real PTA data sets requires $N\gg2$, it is instructive to take a closer look at the $N=2$ scenario, as it allows to obtain some further insights about the likelihood behaviour.

\subsubsection{For two pulsars}

For simplicity, we focus on a single frequency bin. Letting $\vec{\mathcal{A}}=(\mathcal{A}_1, \mathcal{A}_2)$, $\Delta\phi = \phi_2 - \phi_1$,
the integral in Eq.~\ref{eq:phase_llh} reads
\begin{multline}
    p(\vec{\phi}|A_\gwb, \gamma_\gwb) \sim 
        \intl_0^\infty\!\rm{d}\mathcal{A}_1 \rm{d}\mathcal{A}_2\; \mathcal{A}_1\mathcal{A}_2\\ \exp\left[\Sigma^{-1}_{11}\mathcal{A}_1^2 + \Sigma^{-1}_{22}\mathcal{A}_2^2 + 2\Sigma^{-1}_{12}\,\mathcal{A}_1\mathcal{A}_2\,\cos{\Delta\phi} \right].
\end{multline}
Upon using completion of the square and using standard error function identities, we arrive at
\begin{multline}
    p(\vec{\phi}|A_\gwb, \gamma_\gwb) = p(\Delta\phi|A_\gwb, \gamma_\gwb) \sim \\
    \frac{1-r\cos(\Delta\phi)\;\arccos[r\cos(\Delta\phi)]}{\sqrt{1-r^2\cos^2(\Delta\phi)}},
\end{multline}
with $r = \Sigma^{-1}_{12}/\sqrt{\Sigma^{-1}_{11}\Sigma^{-1}_{22}}$.

This shows that the presence of a correlated signal induces a cosine-modulated correlation between the phases of distinct pulsar pairs. This corresponds to the striped signature identified in Fig.~\ref{fig:phasedistribution_example_2p_2b}. Furthermore, it shows that the correlation strength directly reflects the GWB S/N, as 
\begin{equation}
    r = \frac{\Sigma^{-1}_{12}}{\sqrt{\Sigma^{-1}_{11}\Sigma^{-1}_{22}}} = \left(\frac{\Gamma_{12}S_\gwb(f)}{\sigma_1\sigma_2}\right)^{-1}.
\end{equation}
Thus, the correlation pattern is the key to (i) identifying the presence of a GWB signal and (ii) constraining its spectral parameters purely from the phases of the timing residuals, as overall S/N of the correlation between two pulsars depends on the Hellings-Downs coefficient, it scales with the GWB spectrum. 

Note that if the pulsar noise model is misspecified, then the phase likelihood becomes sub-optimal because the prominence of the stripes in the likelihood no longer match the actual distribution of the data.
This can introduce a bias in the inferred values of $A_\text{GWB}, \gamma_\text{GWB}$.
However, even in this case, the phase likelihood is still relatively conservative because $A_\text{GWB}=0$ data \textit{always} produces a uniform phase distribution, which guards against false-positives inferences of $A_\text{GWB}>0$.

\subsubsection{For $N>2$ pulsars}
The general expression for the phase-based likelihood, Eq.~\ref{eq:phase_llh}, is a correlated Gaussian integrated over the positive orthant of the amplitude-space.
The major complication in evaluating this integral arises from the pair-wise correlations between phases, introduced from the non-diagonal $\mathbf{\Sigma}$. Just as in the $N=2$-case, the likelihood is still a function of $\Delta\phi$, however now, $\Delta\phi\longmapsto \vec{\Delta\phi} \equiv\left\{\phi_1-\phi_2,\ldots,  \phi_1-\phi_N, \ldots, \phi_{N-1}-\phi_N\right\}$. These correlations cannot be resolved, meaning that there is no analytic expression for the integral in more than two dimensions, so non-analytical approaches are required.

Evaluating higher-dimensional positive-orthant integrals are a well known problem across many fields, such as probability theory \cite[see]{GenzBretz_2009}, but also economics \citep{Train_2009}, stochastic programming \citep{Prekopa_1973} etc.. For low-dimensional distribtutions, quasi-Monte-Carlo algorithms, e.g.\ introduced by \cite{Genz_1992}, are numerically tractable, in moderate dimensions, the Geweke-Hajivassiliou-Keane \citep{Geweke_1998,Hajivassiliou_1996,Keane_1993} is usually applied. For a typical PTA data set \citep{EPTA_DR2_GWB,NANOGrav15yr_GWB,PPTA_DR3_GWB,MPTA_4.5yr_GWB}, with $\mathcal{O}(30)$ pulsars and $\mathcal{O}(10)$ relevant frequency bins, the integral is taken over $\mathcal{O}(300)$ dimensions. In this regime, all mentioned methods become either numerically unstable or not feasible. While approaches such as sequential Monte-Carlo are proposed \citep{Ridgway_2014} for these dimensions, evaluating Eq.~\ref{eq:phase_llh} using sampling methods would require a nested approach, i.e.\ the integral has to be evaluated at each point of the parameter space. This comes at a high computational cost we aim to avoid in a simple demonstration setup.

Another option is to numerically fit the phase likelihood using simulation-based inference (SBI). 
We explore this option Sec.~\ref{sec:demonstration}.

\subsection{Detection statistic}
\label{ssec:detection_statistic}

Under the null hypothesis, $\mathcal{H}_0$, namely that no GWB signal is present\footnote{For example, the pulsar noise and a CURN signal could be seen as the null-hypothesis.}, $\mathbf{\Sigma}$ is (block-)diagonal, therefore $\mathbf{M}(\vec{\phi})$ is diagonal as well. The corresponding phase-based likelihood is uniform,
\begin{equation}
    \mathcal{L}(\mathcal{H}_0 |\vec{\phi}) = \frac{1}{(2\pi)^N}.
\end{equation}

The detection statistic used to test for the presence of a GWB signal (signal hypothesis, $\mathcal{H}_1$), is the likelihood ratio, which is given as
\begin{align}
    \Lambda &= \frac{\mathcal{L}(A_\gwb, \gamma_\gwb|\vec{\phi})}{ \mathcal{L}(\mathcal{H}_0 |\vec{\phi})} \\
            &= (2\pi)^N\cdot \mathcal{L}(A_\gwb, \gamma_\gwb|\vec{\phi}).
\end{align}

The null distribution of the likelihood ratio is a fixed function of a uniform vector. It can be computed without touching the data, by drawing realisations of $\vec{\phi}_0\sim \mathrm{Uniform([-\pi,\pi]^N})$ and evaluating $\Lambda_0 = (2\pi)^N\cdot \mathcal{L}(A_\gwb, \gamma_\gwb|\vec{\phi}_0)$. Therefore, the detection statistic satisfies criteria (1) and (2) listed in Sec.~\ref{sec:intro}. The corresponding Bayes factor comparing signal and noise hypotheses is obtained by marginalising over the GWB parameters, $A_\gwb$ and $\gamma_\gwb$.

Note that these considerations assume that the timing residuals are not contaminated by monopolar or dipolar common noise, e.g.\ arising from clock noise or Solar system ephemeris noise. These noise processes would create a non-uniform null distribution, since they would also lead to off-diagonal terms in $\mathbf{\Sigma}$.

\section{Demonstration}
\label{sec:demonstration}
We create a mock PTA consisting of $N_\mathrm{psr}=30$ pulsars at randomly drawn sky positions. We assume that the pulsars have been regularly observed over 15 years, and that their residuals contain only white noise with $\sigma_\mathrm{ToA}=\SI{50}{\nano\second}$ and EFAC$=1$, and a GWB signal. We model the GWB over the first 12 frequency bins. Furthermore, we assume the white noise level to be known and fixed, such that in this mock example, the phase likelihood only depends on the amplitude and spectral index of the GWB.

Since numerically evaluating the likelihood for $(A_\gwb, \gamma_\gwb)$ obtained from Eq.~\ref{eq:phase_llh} in a Monte Carlo approach is challenging, we pursue simulation-based inference for exploring the toy model. This means, that we use forward modelling and generate an empirical distribution of ${\cal L}( A_\gwb, \gamma_\gwb | \vec{\phi})$, which is then fitted using a conditional normalising flow. For readers unfamiliar with the concepts of normalising flows, we provide a brief overview in App.~\ref{app:nf}.

\subsection{Setup and training of the normalising flow}

The phases of complex numbers are circular over $[-\pi,\pi)$, i.e., 
\begin{align}
    {\cal L}(\phi=-\pi) =  
    {\cal L}(\phi = \pi) .
\end{align}
In order to properly account for this characteristic, we choose to model the likelihood using a neural circular spline flow (NCSF) \citep{Rezende_2020}. This special type of neural spline flow \citep{Durkan_2019} respects the circular nature of complex phases, by ensuring that the splines at $\{-\pi, \pi\}$ are continuous up to their first derivative. A spline flow uses monotonic quadratic-rational spline functions as the bijection. The parameters (widths, heights, and derivatives) of these splines at each data point are determined by a neural network, that is conditioned during the training process. This setup is especially suited to model PDFs with the numerical stability needed for reliable parameter inference. We employ the implementation of such a flow provided in the \textsc{Python} package \textsc{zuko}.\footnote{\url{https://pypi.org/project/zuko/}}

Since the distribution of phases depends on the gravitational-wave parameters $A_\text{GWB}, \gamma_\text{GWB}$, we need to train a \textit{conditional flow} describing the distribution of the phases conditioned on these parameters.
A conditional flow is trained not only on the phases, but also on the parameters these phases were generated with, such that it directly represents ${\cal L}(\vec{\phi}| A_\gwb, \gamma_\gwb)$. 
Flows are especially effective when the likelihood varies smoothly as function of the the input parameters.

While for real PTA data sets the irregular sampling of the timing residuals leads to non-negligible correlations between frequency bins, due to the regular-sampling assumption made in the simulation, we can neglect these correlations for the test case. This decouples the likelihood in frequency, such that
\begin{equation}\label{eq:approx_llh}
    \mathcal{L}_\mathrm{approx}( A_\gwb, \gamma_\gwb | \vec{\phi}) = \prod_{k=1}^\mathrm{N_{f}} \mathcal{L}( A_\gwb, \gamma_\gwb, f_k | \vec{\phi}).
\end{equation}
This decoupling allows us to reduce the complexity of the flow. Instead of modelling the full $(N_\mathrm{PSR}\times N_\mathrm{bin})$-dimensional likelihood, we can fit each bin separately and then combine the flows.
It may be possible to fit every frequency bin simultaneously, but we leave this for future work.
 
We empirically find that a conditional NCSF with 15 layers, each consisting of an 8-bin spline and a multi-layer perceptron \citep{RumelhartHintonWilliams_1986} with two layers with $[64, 64]$ features, provides an adequate fit. 
We train this flow in multiple training rounds.
Per round, we draw $N_\mathrm{vp}$ pairs of $(\log A_\gwb, \gamma_\gwb)$ according to a Sobol sequence, which ensures a more uniform coverage of the sample space than random sampling, leading to faster convergence and lower variance estimates while avoiding gridding artefacts in the flow training arising from regular parameter spacing. We use a validation fraction of $0.1$, i.e., 10\% of these points are randomly chosen to represent the validation data set, whereas the remaining 90\% points make up the training data set. At each of these points, we draw $N_\mathrm{sample}$ phase samples. In a single round, the flow is then trained in multiple epochs using the \textsc{AdamW} optimizer \citep{LoshchilovHutter_2018} provided by the \textsc{Python} package \textsc{PyTorch}, with an initial learning rate, $\eta_0$ that is adapted for each round. Each round is finished, if the relative change in the validation NLL does not exceed a manually set threshold for more than five epochs. The specific training parameters are summarised in Tab.~\ref{tab:train_params}. We verify that the flow adequately represents the phase distribution using a $P$-$P$-plot, shown in App.~\ref{app:freq_fac_flow_checks}. 

\begin{table}[htbp]
    \centering
    \begin{tabular}{l|l l l l}
        Round   & $N_\mathrm{vp}$   & $N_\mathrm{sample}$   & $\eta_0$    & rel. change   \\ \hline
        1       & 500               & 15360                 & \num{1e-3}  & \num{5e-3}    \\
        2       & 1000              & 15360                 & \num{3e-4}  & \num{1e-4}    \\
        3       & 1000              & 30720                 & \num{1e-4}  & \num{5e-5}
    \end{tabular}
    \caption{Training parameters for the 30-dimensional conditional normalising flow. The flow is a neural circular spline flow with 15 layers.}
    \label{tab:train_params}
\end{table}

\subsection{Evaluating amplitude and spectral index}

\begin{figure}
    \centering
    \includegraphics[width=\linewidth]{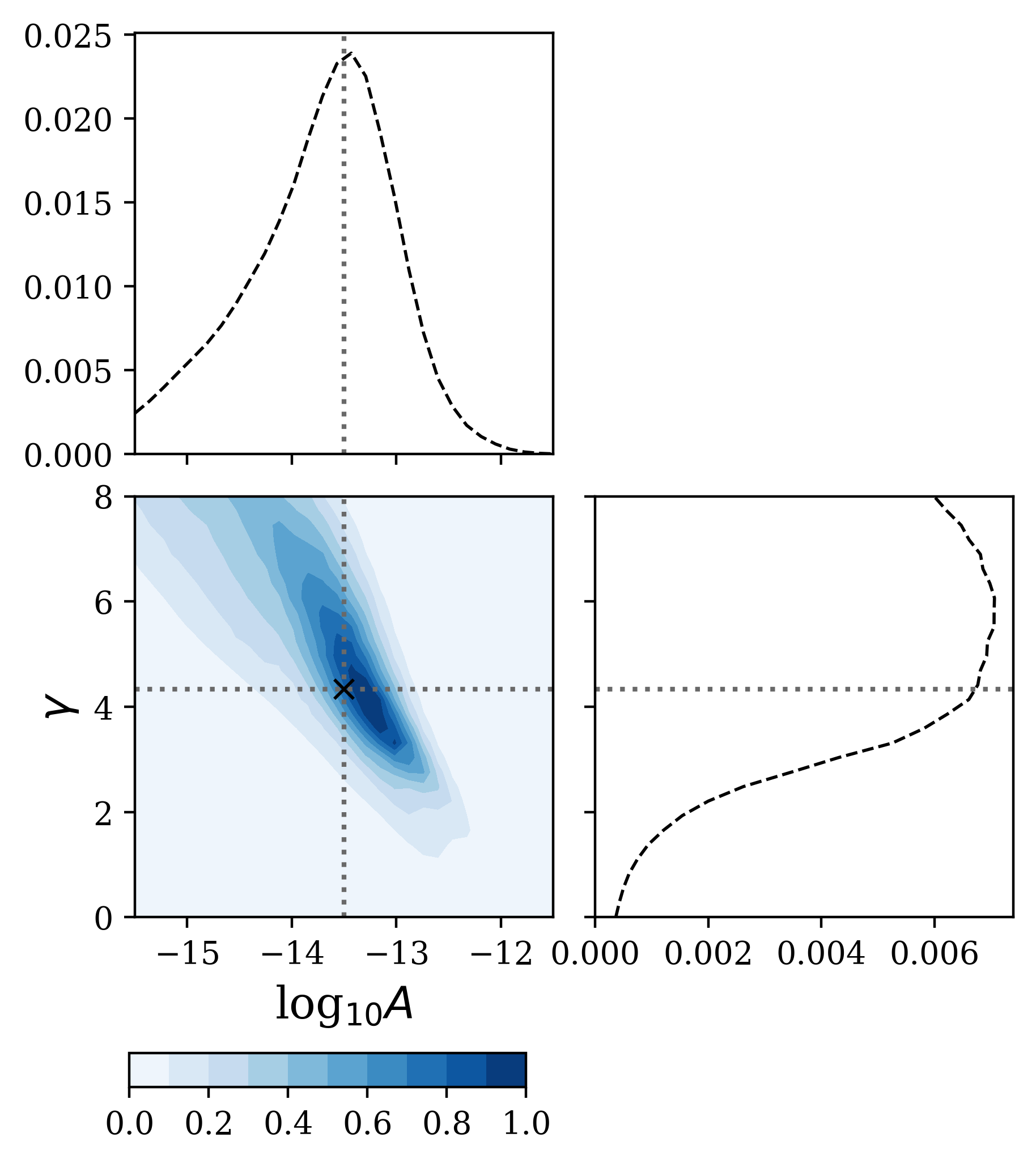}
    \caption{The phase likelihood as fit with a neural circular spine flow.
    This example includes 30 pulsars and 12 frequency bins. The cross and the dashed lines indicate the injected GWB signal, with $\log_{10}\!A_\gwb^\mathrm{inj}=-13.5$ and $\gamma_\gwb^\mathrm{inj} = 13/3$.}
    \label{fig:llh_A_gamma_30psrFRES}
\end{figure}

We now demonstrate the capability of the phase likelihood to determine the presence of a gravitational-wave background with parameters, $\log_{10}\!A_\gwb$ and $\gamma_\gwb$. To this end, we create simulated data with a gravitational-wave background signal described by parameters $\log_{10}\!A_\gwb^\mathrm{inj}=-14.0$ and $\gamma_\gwb^\mathrm{inj} = 13/3$ and draw a random realisation of complex residuals, $\vec{\mathcal{Z}}^\mathrm{inj}$. Then we obtain the phases of these residuals, and calculate the likelihood by evaluating the trained flow for these phase values across the $(\log_{10}\!A_\gwb, \gamma_\gwb)$-plane according to Eq.~\ref{eq:approx_llh}. The resulting likelihood contour for a single realisation of phases is shown in Fig.~\ref{fig:llh_A_gamma_30psrFRES}. 
We find that the likelihood peak is consistent with the injected values of $(\log_{10} A_\text{GWB}, \gamma_\text{GWB})$. 
Since the confidence interval excludes small values of $A_\text{GWB}$, we conclude that it is possible to detect and recover a GWB signal using purely the phase information of the timing residuals.

It is useful to think about how $\log_{10} A_\text{GWB}, \gamma_\text{GWB}$ are imprinted on the phase likelihood.
The spectral index determines how the intensity of the stripes varies with frequency. When $\gamma_\text{GWB}$ is large, the prominence of the stripes falls relatively faster as a function of frequency whereas small values values of $\gamma_\text{GWB}$ make it easier to see the stripes at higher frequencies.
Increasing $A_\text{GWB}$ increases the visibility of the stripes in all frequency bins---at least up to a point. 
However, the two parameters are somewhat covariant.
In the high-SNR limit (when the gravitational-wave background is large compared to the noise) the phases of two pulsars are still not perfectly correlated because the gravitational-wave background enters as a source of noise via the pulsar term.

\subsection{Evaluating the statistical significance}

\begin{figure}
    \centering
    \includegraphics[width=\linewidth, trim={0 0 0 0},clip]{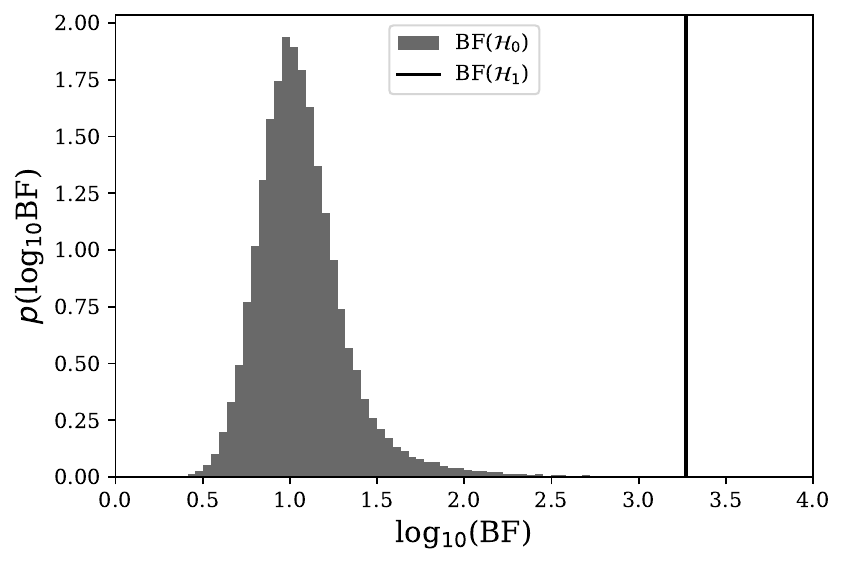}
    \caption{Distribution of Bayes factors in the absence of a GWB signal (null-hypothesis, $\mathcal{H}_0$), and the Bayes factor for the likelihood of the phases containing a GWB signal, with $\log_{10}\!A_\gwb^\mathrm{inj}=-13.5$ and $\gamma_\gwb^\mathrm{inj} = 13/3$ (signal hypothesis, $\mathcal{H}_1$),. The $p$-value of the observed Bayes factor is $0.00047$, corresponding to a statistical significance of $3.3\,\sigma$.}
    \label{fig:bayesfactor}
\end{figure}

As described in Sec.~\ref{ssec:detection_statistic}, the distribution of phase vectors under the null hypothesis, $\mathcal{H}_0$, for the 30-pulsar PTA evaluated in 12 frequency bins is given as $\vec{\phi}_0\sim \mathrm{Uniform([-\pi,\pi]^{30\times12}})$. We generate the distribution of null-hypothesis Bayes factors from 100000 samples of $\vec{\phi}_0$. For each sample, we calculate the corresponding Bayes factors by integrating the respective likelihood obtained from the trained neural circular spline flow over the $\log_{10}A_\gwb$-$\gamma_\gwb$ parameter space. Similarly, we calculate the Bayes factor of the phase vector obtained under the signal hypothesis, $\mathcal{H}_1$, which was used to obtain Fig.~\ref{fig:llh_A_gamma_30psrFRES}. The resulting distribution of $\{\log_{10}\mathrm{BF}(\mathcal{H}_0)\}$ and $\log_{10}\mathrm{BF}(\mathcal{H}_1)$ are shown in Fig.~\ref{fig:bayesfactor}. The Bayes factor observed from the phases containing the GWB signal, $\mathrm{BF}(\mathcal{H}_1$), has a $p$-value of $0.00047$, which corresponds to a statistical significance of $3.3\,\sigma$. This evaluation demonstrates the procedure of obtaining the statistical significance for the presence of a GWB signal without resampling a likelihood obtained from the observed data.

\subsection{Caveats}

This demonstration features a simplified representation of a PTA data set. Realistic timing residual time series are irregularly sampled, obtained at different observing frequencies and have different overall time spans. This leads to correlations between frequency bins that are not reflected in this simulation. In Appendix~\ref{app:freq_res_flow} we attempt to overcome this by modelling the joint $(N_\mathrm{psr}\times N_\mathrm{bin})$-dimensional likelihood.

Furthermore, as the number of dimensions increases (either from adding more frequency bins or more pulsars) modelling the likelihood using a normalising flow becomes increasingly challenging---at least when employing an off-the-shelf model such as the pre-implemented NCSF. 

Modelling the likelihood in order to constrain only the GWB amplitude proved to be less dependent on the specific flow architecture, and once we found a suitable setup for a small number of pulsars, this was easily scalable, both in the construction of the flow as well as its training. However, determining a setup (flow + training) that is capable of modelling the dependencies of the likelihood conditioned on both the amplitude and spectral index proved to be hard, already for a fairly simple model. 

Information about the GWB is mostly encoded in subtle correlations between specific pairs of pulsars.
It is difficult for the normalising flow to learn how this subtle structure varies with $A_\text{GWB}, \gamma_\text{GWB}$.
Thus, the setup needs to be fine-tuned to the particular properties of the PTA data set; e.g., adding or removing five pulsars can require alterations to the flow architecture.

\subsection{Comparison to the Optimal Statistic}
We compare the performance of the phase-based detection statistic to the currently used detection statistic, the optimal statistic (OS), using the framework described in \cite{vanHaasteren_2025p}. As shown in Appendix~\ref{app:pcos}, the OS is a quadratic function of the residuals, combined by a filter matrix, $\mathbf{Q}$, which, in Fourier domain, is given as
\begin{equation}
    D(\vec{\mathcal{Z}}, \theta) = \vec{\mathcal{Z}}^\dagger\mathbf{Q}\vec{\mathcal{Z}},
\end{equation}
as a function of the model parameters, $\theta$.
The filter matrix,
\begin{equation}
    \mathbf{Q} = \frac{\mathbf{N}^{-1}(\mathbf{C}-\mathbf{N})\mathbf{N}^{-1}}{\mathrm{Tr}\left[\left(\mathbf{N}^{-1}(\mathbf{C}-\mathbf{N})\right)^2\right]^{1/2}},
\end{equation}
is constructed from the covariance matrices of the timing residuals in the absence and presence of a GWB signal,
\begin{align}
    \mathbf{N} &= \langle\vec{\mathcal{Z}}\vec{\mathcal{Z}}^\dagger\rangle_{H_0} = \mathrm{diag}(\mathbf{\Sigma})\\
    \mathbf{C} &= \langle\vec{\mathcal{Z}}\vec{\mathcal{Z}}^\dagger\rangle_{H_S} = \mathbf{\Sigma}.
\end{align}

\begin{figure}
    \centering
    \includegraphics[width=\linewidth]{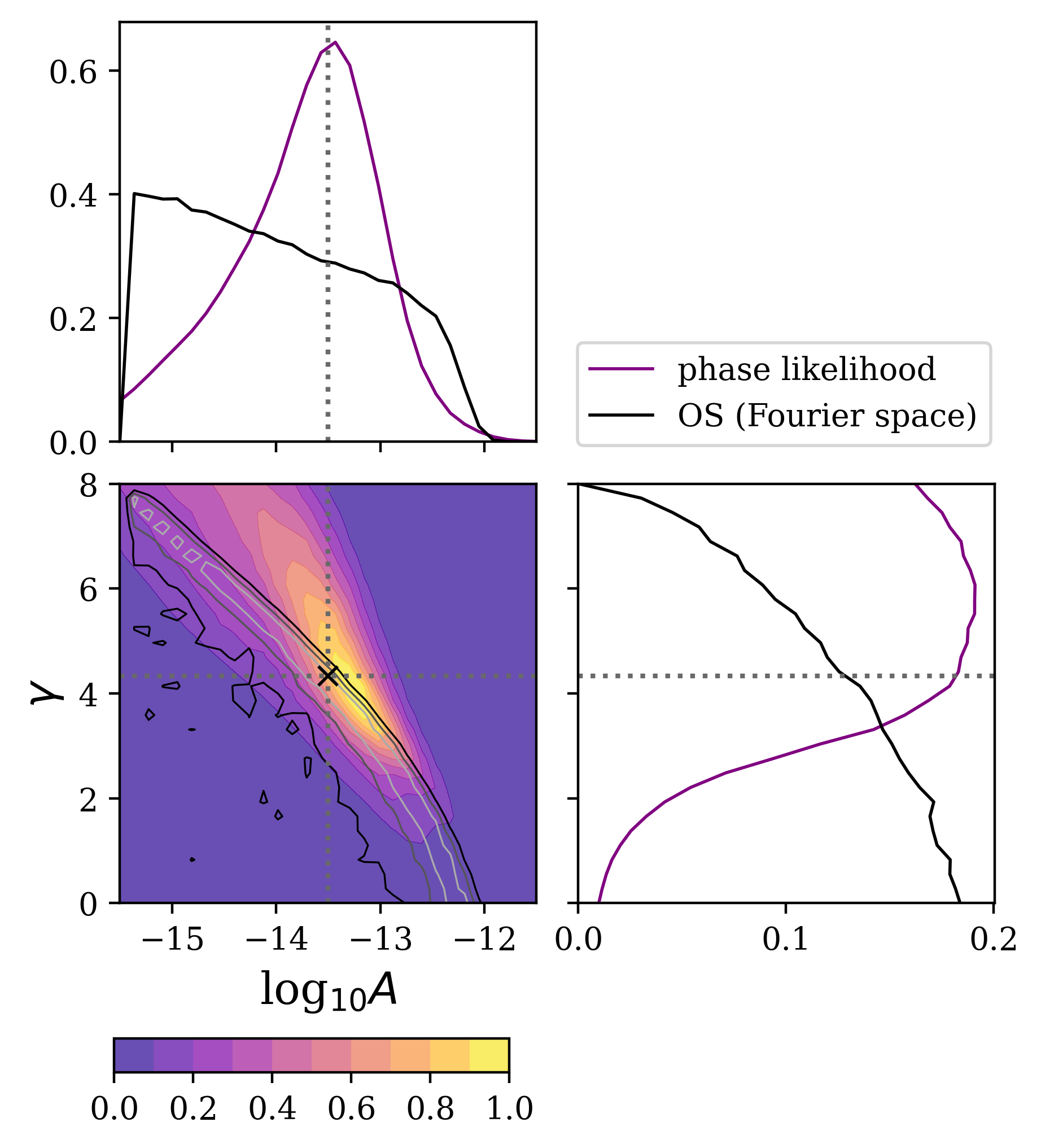}
    \caption{Comparison between GWB parameter likelihood from evaluating the phase likelihood and the Optimal Statistic distribution from a single draw of timing residuals. The cross and the dashed lines indicate the injected GWB signal, with $\log_{10}\!A_\gwb^\mathrm{inj}=-13.5$ and $\gamma_\gwb^\mathrm{inj} = 13/3$.}
    \label{fig:likelihood_comp}
\end{figure}

We evaluate the likelihood for the GWB parameters, $\theta = (\log_{10}A_\gwb, \gamma_\gwb)$, by generating the empirical distribution of $D(\vec{\mathcal{Z}}, \theta)$ for different value pairs of $\theta$, and evaluating these for a randomly drawn residual sample, $\vec{\mathcal{Z}}_0$. The resulting parameter posteriors obtained from the WN+GWB-only, 30-pulsar PTA test case are shown in Fig.~\ref{fig:likelihood_comp}, together with the contours of the phase-based likelihood evaluation from Fig.~\ref{fig:llh_A_gamma_30psrFRES}.

\begin{figure}
    \centering
    \includegraphics[width=\linewidth]{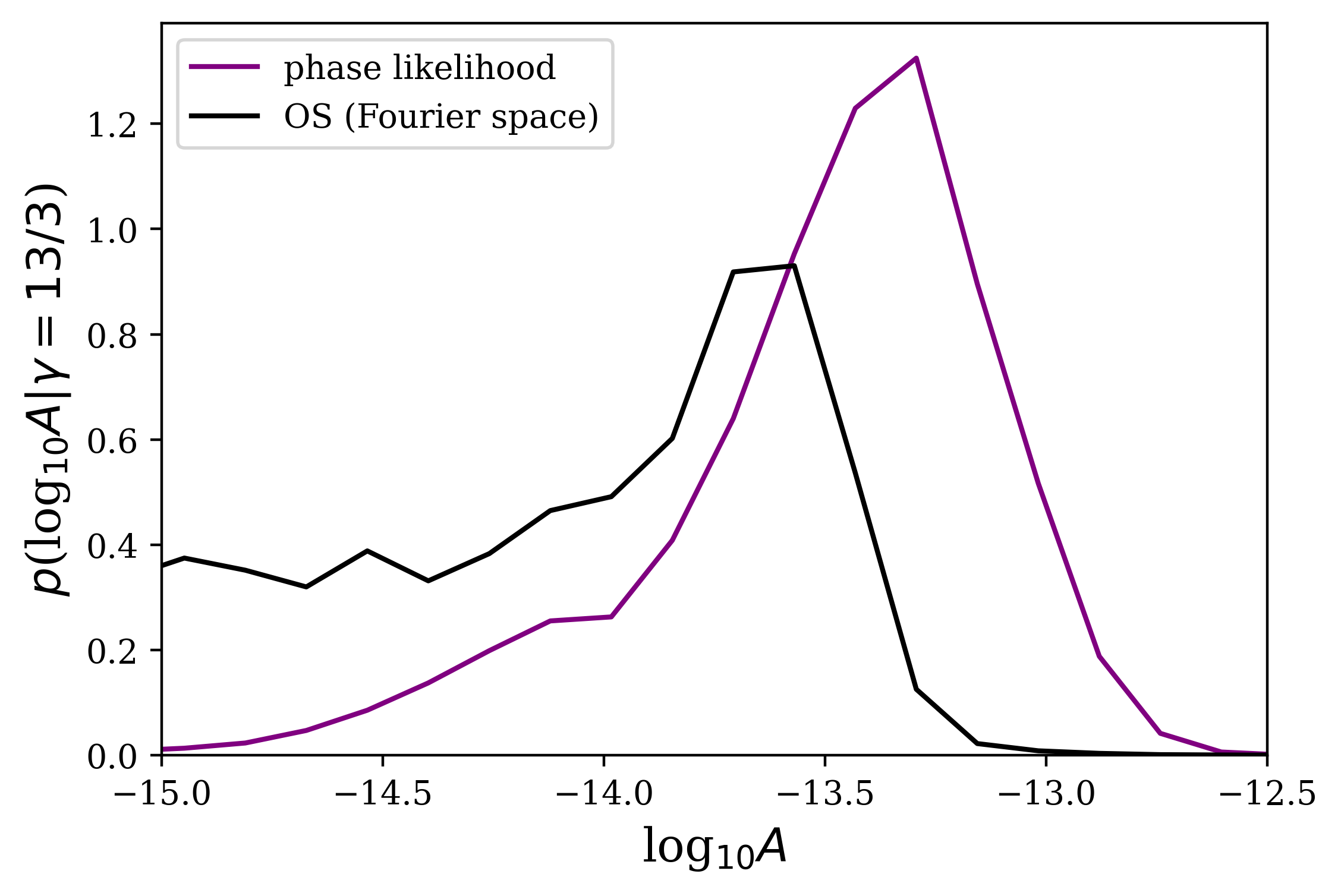}
    \caption{One-dimensional slice of the two-dimensional posteriors shown in Fig.~\ref{fig:likelihood_comp} for $\gamma_\gwb=13/3$.}
    \label{fig:likelihood_slice}
\end{figure}

Fig.~\ref{fig:likelihood_comp} solidifies that the phase-based likelihood can be used to detect GW signals. While it provides slightly wider 2D contours than the OS, it performs better in separating $\log_{10}A_\gwb$ and $\gamma_\gwb$. This narrower posterior from the OS corresponds to a tighter constraint on the GWB amplitude obtained from the OS, once the spectral index has been fixed, as shown in Fig.~\ref{fig:likelihood_slice}. However, this suggests, that the OS estimate likely includes some amplitude ($\{\mathcal{A}_i\}$) information, which the community has indicated to actually not trust.

\section{Summary and outlook}
\label{sec:summary_outlook}

In this work, we construct a likelihood function for the phases of the complex residuals of a pulsar timing array. 
A gravitational-wave signal appears in the phase data as correlations between the phases of different pulsars.
We use this likelihood to create a formalism with the three properties laid out in the introduction: 
\begin{enumerate}
    \item It is safe from false positives when the pulsar noise model is misspecified.
    \item It does not require quasi-resampling methods to assess significance.
    \item It makes the optimal use of the (phase) information we trust in order to constrain the properties of the gravitational-wave background.
\end{enumerate}
Most importantly, we demonstrate that the null distribution of the phase-based likelihood does not depend on the choice of noise parameters.
By comparing this framework to the traditional optimal statistic approach (which does depend on the choice of noise parameters), we show that the optimal statistic still uses some information about the pulsar amplitudes in order to resolve the presence of a gravitational-wave background---in spite of community concerns about the reliability of pulsar noise modelling.

That said, we view this work as a proof of concept; we have not produced a mature pipeline capable of analysing a complete PTA dataset with a large number of pulsars and frequency bins.
In order to elevate this approach beyond a conceptual idea towards a framework, several hurdles have to be overcome.
First, additional work is required to account for the covariance between different frequency bins. This goes hand-in-hand with an increase in the dimensionality of the likelihood, both by adding more frequency bins and/or more pulsars. So, second, it is probably necessary to employ a more sophisticated normalising flow architecture.
Third, in order to take into account the uncertainty in our pulsar noise models in the parameter estimation, one should include random draws from pulsar noise parameters in our simulation-based inference. \\
Another direction of development is the type of signal that is evaluated. At the time of writing, the phase likelihood can be used to assess the significance of the isotropic background signal with only phase information. As a result of future work, it can potentially be used to assess the phase-only significance of other signals such as continuous-wave signals.\\

In the closing remarks to this work, we want to address the provocative question that is looming over any future work: should we fully turn to a phase likelihood for pulsar-timing inference?
We have a nuanced answer to this question.
If the choice is between an optimal statistic framework (built with cross-correlations) and a phase likelihood, we submit that the phase likelihood is preferable because it actually removes the amplitude information that the optimal statistic is meant to ignore (but does not). 
As a result, the community can avoid the need for quasi-resampling tests to measure significance; the Bayes factors obtained from the phase-based likelihood can be taken at face value. \\
In the long term however, it might not make sense to completely throw out all amplitude information. As the field of PTAs moves toward what is likely to be a ``post-detection'' era, the conservatism to avoid a false-positive detection (in the intermediate signal-to-noise regime) may be relaxed.
At that point, the idea of removing all amplitude information seems unnecessary, and other methods may become relevant\footnote{This is comparable to the situation of terrestrial observatories like LIGO and Virgo. They also deal with misspecified noise, but they do not throw out information---they just do their best to model the noise \citep{RomeroShaw_2022}}.
Overall, we see the field as eventually moving away from cross-correlation toward the use of pulsar amplitudes and phase. In this sense, we think the phase likelihood will become useful for determining where the information on a GW signal is coming from. Therefore, we envision the phase likelihood as providing a valuable tool for determining the extent to which our inferences depend on less reliable amplitude measurements.

\section*{Acknowledgements}

KG, DC and MK acknowledge continuing valuable support from the Max-Planck society. KG acknowledges support from the International Max Planck Research School (IMPRS) for Astronomy and Astrophysics at the Universities of Bonn and Cologne.
ET acknowledges support from the Australian Research Council (ARC) Centres of Excellence CE230100016 and ARC DP230103088.
MK acknowlegdes support by the CAS-MPG legacy Programme.

This publication made use of open source python libraries including \textsc{numpy} \citep{numpy}, \textsc{scipy} \citep{scipy}, \textsc{matplotlib} \citep{matplotlib}, \textsc{PyTorch} \citep{PyTorch} and \textsc{zuko} (\url{https://pypi.org/project/zuko/}).
The authors acknowledge the use of the large-language-models ChatGPT and Claude to assist in code-writing.

\bibliographystyle{aa}
\bibliography{refs}

\appendix

\section{Summary of the pair-correlation optimal statistic}
\label{app:pcos}

The pair-correlation optimal statistic (OS) \citep{Demorest_2013,Chamberlin_2015} is built from pulsar-pair cross-correlation estimators, $\rho_{ab}$, where $(a,b)$ indicate distinct pulsars. These estimators, and their uncertainties, $\sigma_{ab}$, are obtained from the timing residuals, $\vec{\toa}$, as
\begin{align}\label{eq:tech:broadband_cc_est}
    \rho_{ab} &= \frac{\vec{\toa}_a^T\; \mtx{C}_a^{-1} \hat{\mtx{S}}_{ab}\mtx{C}_b^{-1}\;\vec{\toa}_b}{\mathrm{Tr}\left( C_a^{-1}\hat{\mtx{S}}_{ab} C_b^{-1}\hat{\mtx{S}}_{ba}\right)}, \\
    \sigma_{ab} &= \left[\mathrm{Tr}\left( C_a^{-1}\hat{\mtx{S}}_{ab} C_b^{-1}\hat{\mtx{S}}_{ba}\right)\right]^{-1/2},
\end{align}
where $C_a = \langle\vec{\toa}_a\vec{\toa}_a^T\rangle$ refers to the autocorrelation matrix of the timing residuals, and $S_{ab}$
is the cross-correlation covariance as introduced in Eq.~\eqref{eq:residuals_ccmatrix}. 
With these quantities, the estimator for the GWB amplitude is given as
\begin{equation}
    \hat{A}^2_\gwb = \frac{\sum_{a<b}\rho_{ab}\Gamma_{ab}/ \sigma_{ab}^2}{\sum_{a<b}\Gamma_{ab}^2/\sigma_{ab}^2},
\end{equation}
and follows $\langle\hat{A}^2\rangle  =A^2_\gwb$. The signal-to-noise of the amplitude estimator, 
\begin{equation}
     \hat{\rho}= \frac{\sum_{a<b}\rho_{ab}\Gamma_{ab}/ \sigma_{ab}^2}{\left[\sum_{a<b}\Gamma_{ab}^2/\sigma_{ab}^2\right]^{1/2}},
\end{equation}
is the frequentist detection statistic, which is widely used in PTA GWB analyses \citep{NANOGrav15yr_GWB,EPTA_DR2_GWB,PPTA_DR3_GWB,MPTA_4.5yr_GWB}.

\section{Basics of Normalising Flows}
\label{app:nf}

A normalising flow is a transformation built from a sequence of differentiable bijections, which converts a simple base probability distribution, e.g., a normal distribution, into a more complex target distribution. On the one hand, this allows to sample from the target distribution by transforming samples from the base distribution. On the other hand, it enables an efficient evaluation of the density at a sample (meaning the likelihood) by transforming the sample to the probability space of the base distribution. Therefore, normalising flows are a popular tool for variational inference and density estimation. 

Assuming that the random variable, $\vec{X}\in\mathbb{R}^D$ with known probability density function $p_{\vec{X}}(\vec{x})$ is related to another random variable $\vec{Y}=\vec{f}(\vec{X})$ via the bijection $g$, the probability density function of $\vec{Y}$, $p_{\vec{Y}}(\vec{y})$ can be obtained as
\begin{align}
    p_{\vec{Y}}(\vec{y}) &= p_{\vec{X}}(\vec{x}) \left|\det\left[\frac{\partial \vec{x}}{\partial\vec{y}}\right] \right| \\
    &=p_{\vec{X}}\left(\vec{f}(\vec{y})\right) \left|\det\left[\frac{\partial \vec{f}(\vec{y})}{\partial\vec{y}}\right] \right|,
\end{align}
where $\vec{f}$ is the inverse of $\vec{g}$, and $\frac{\partial \vec{f}}{\partial\vec{y}}$ is the Jacobian of $\vec{f}$, which accounts for the volume change of the probability space due to the transformation.
The obtained density function $p_{\vec{Y}}(\vec{y})$ is referred to as the pushforward of $p_{\vec{X}}(\vec{x})$.

With regards to generative models, the function $\vec{g}$ pushes forward the simple base distribution to a more complicated density. This direction is called the `generative direction', and $\vec{g}$ is the `generator'. The inverse function, $\vec{f}$, moves, or `flows', the more complicated density back to the base distribution. This is the `normalising direction', as $\vec{f}$ `normalises' the data distribution. This leads to the term `normalising flow'.

In general, it is possible to construct arbitrarily complex (non-linear) bijections, such that any target distribution can be generated from the base distribution in a single transformation. However, in constructing normalising flows, it is desirable, that the bijection is sufficiently feasible to evaluate and invert, and that the determinant of its Jacobian can be calculated with moderate computational expense. To this end, Normalising Flows exploit that the composition of $N$ bijections,
\begin{equation}
    \vec{g} = \vec{g}_N\,\circ\,\vec{g}_{N-1}\,\circ \ldots  \,\circ\,\vec{g}_1,
\end{equation}
is bijective itself, with inverse
\begin{equation}
    \vec{f} = \vec{f}_1\,\circ\, \ldots \,\circ \vec{f}_{N-1}\,\circ\,\vec{f}_N.
\end{equation}
Furthermore, the determinant of the Jacobian factorises as
\begin{equation}
    \det\left[\frac{\partial \vec{f}(\vec{y})}{\partial\vec{y}}\right] = \prod_{i=1}^N\det\left[\frac{\partial \vec{f}_i(\vec{y}_i)}{\partial\vec{y}_i}\right].
\end{equation}
The individual bijections, or repeating groups of bijections are called the `layers' of a normalising flow. The specific choice of the bijections and the number of layers are often referred to as the architecture of the flow.

Initially, the free parameters of the bijections are arbitrarily chosen, and the pushforward of the base distribution does not model the actual target distribution. Thus, these parameters have to be adjusted by iteratively fitting the transformed base distribution to a sufficiently large sample from he true target distribution. This process is known as `training' the flow. The goal of the training is minimising the loss function, which defines how well the flow fits the actual target distribution. The primary loss function used to train a flow is the negative log-likelihood (NLL) of samples from the true distribution, 
\begin{align}
    \mathcal{L}_\mathrm{NF}(\vec{y}_\mathrm{true}) &\equiv -\log\left[p_{\vec{Y}}(\vec{y}_\mathrm{true})\right] \\
    &= -\log\left[p_{\vec{X}}\left((\vec{f}(\vec{y}_\mathrm{true})\right)\,\cdot\,\left|\det\frac{\partial\vec{f}}{\partial\vec{y}}\right|_{\vec{y}_\mathrm{true}} \right].
\end{align}
Minimising this likelihood enforces that the flow produces samples that are distributed according to the original distribution. Therefore, a properly training the flow requires generating a representative sample data set from the target distribution. This data set has to be large enough such that it reflects the relevant features of the distribution.

The following training of the flow is typically performed using an iterative optimisation scheme based on gradient descent. Gradient descent is an algorithm used to minimise a differentiable function, $h(\vec{z})$, by moving from some point in the parameter space, $\vec{z}_n$, against the gradient of the function to the next point, $\vec{z}_{n+1}$, such that
\begin{equation}
    \vec{z}_{n+1} = \vec{z}_n - \eta\, \vec{\nabla}h(\vec{z}_n).
\end{equation}
The learning rate, $\eta$, governs the step size. For suitably chosen values of $\eta$, $h(\vec{z}_{n+1})\leq h(\vec{z}_n)$, and the minimum of $h(\vec{z})$ can be found. One variation of this algorithm is the stochastic gradient descent (SGD), which is often applied in machine learning. It minimises the loss function by updating the model parameters using random subsets (`batches') of the data, instead of the full data set. This introduces noise fluctuations in the optimisation path, however, significantly speeds up the training of the flow for large data sets.

The SGD itself does not provide a criterion to determine whether the normalising-flow parameters are sufficiently trained in order to properly model the target distribution. A commonly used method to evaluate the convergence behaviour is to split the sample data set into a (larger) training data set and a (smaller) validation data set. The percentage of sample points making up the latter is the `validation fraction'. The training data set is used during the SGD. After iterating batch-wise though all training samples, the average NLL across the validation dataset, $\bar{\mathcal{L}}_\mathrm{NF}(\{\vec{y}_\mathrm{valid}\}) \equiv \bar{\mathcal{L}}_\mathrm{NF}^\mathrm{valid}$, is evaluated. Together, the cycle of SGD and calculation of $\bar{\mathcal{L}}_\mathrm{NF}^\mathrm{valid}$ is then repeated, until the relative change of $\bar{\mathcal{L}}_\mathrm{NF}^\mathrm{valid}$ from one iteration to the next one is less than a suitably chosen threshold. Due to the NLL fluctuations appearing during an SGD, it is customary to introduce a latency period before it is assumed that the flow is trained, i.e.\ the relative change of $\bar{\mathcal{L}}_\mathrm{NF}^\mathrm{valid}$ has to be smaller than the imposed threshold for $n_\mathrm{latency}$ sequential cycles. 

Overall, the performance of the normalising flow in modelling the target distribution is determined by the flow architecture, e.g., the choice of bijections and layers, but also the training setup, in detail, the choice of the sample size, the batch size, the validation fraction, the step size and the latency period. The more complicated or higher dimensional a problem, the more fine-tuning of these parameters is required.

\section{Detailed cross-checks for the frequency-factorised normalising flow}
\label{app:freq_fac_flow_checks}

\begin{figure}
    \centering
    \includegraphics[width=\linewidth]{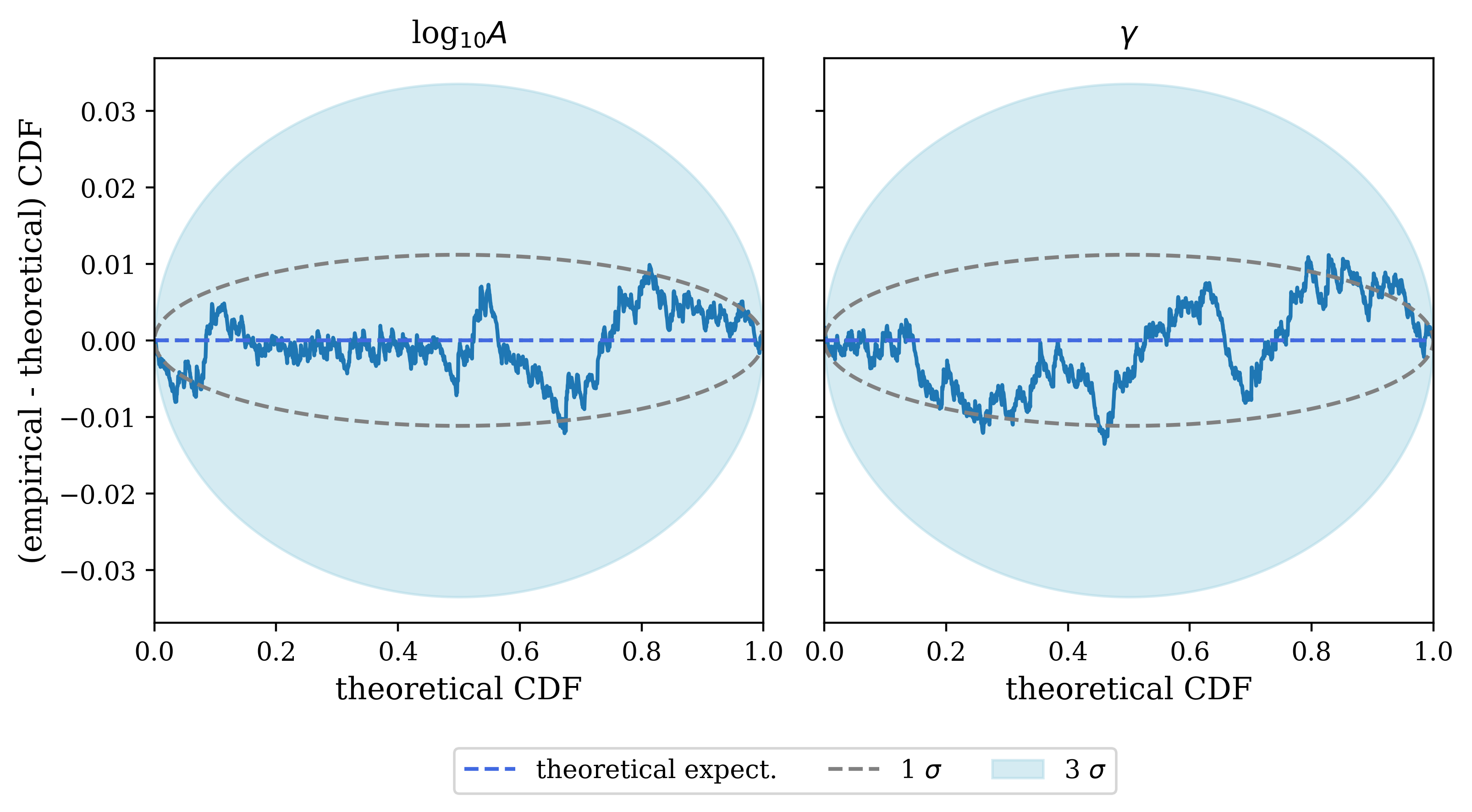}
    \caption{A $P-P$ plot of the recovered log-amplitude (left) and spectral index (right) recovered from 1000 evaluations of the trained conditional neural circular spline flow with frequency dependence. }
    \label{fig:p-p-plot}
\end{figure}

We validate that the trained flow properly represents the empirical likelihood by means of comparing the empirical cumulative density functions for both $\log_{10}\!A_\gwb$ and $\gamma_\gwb$. The resulting $P-P$ plots obtained from 1000 phase samples are displayed in Fig.~\ref{fig:p-p-plot}.

\begin{figure}
    \centering
    \includegraphics[width=\linewidth]{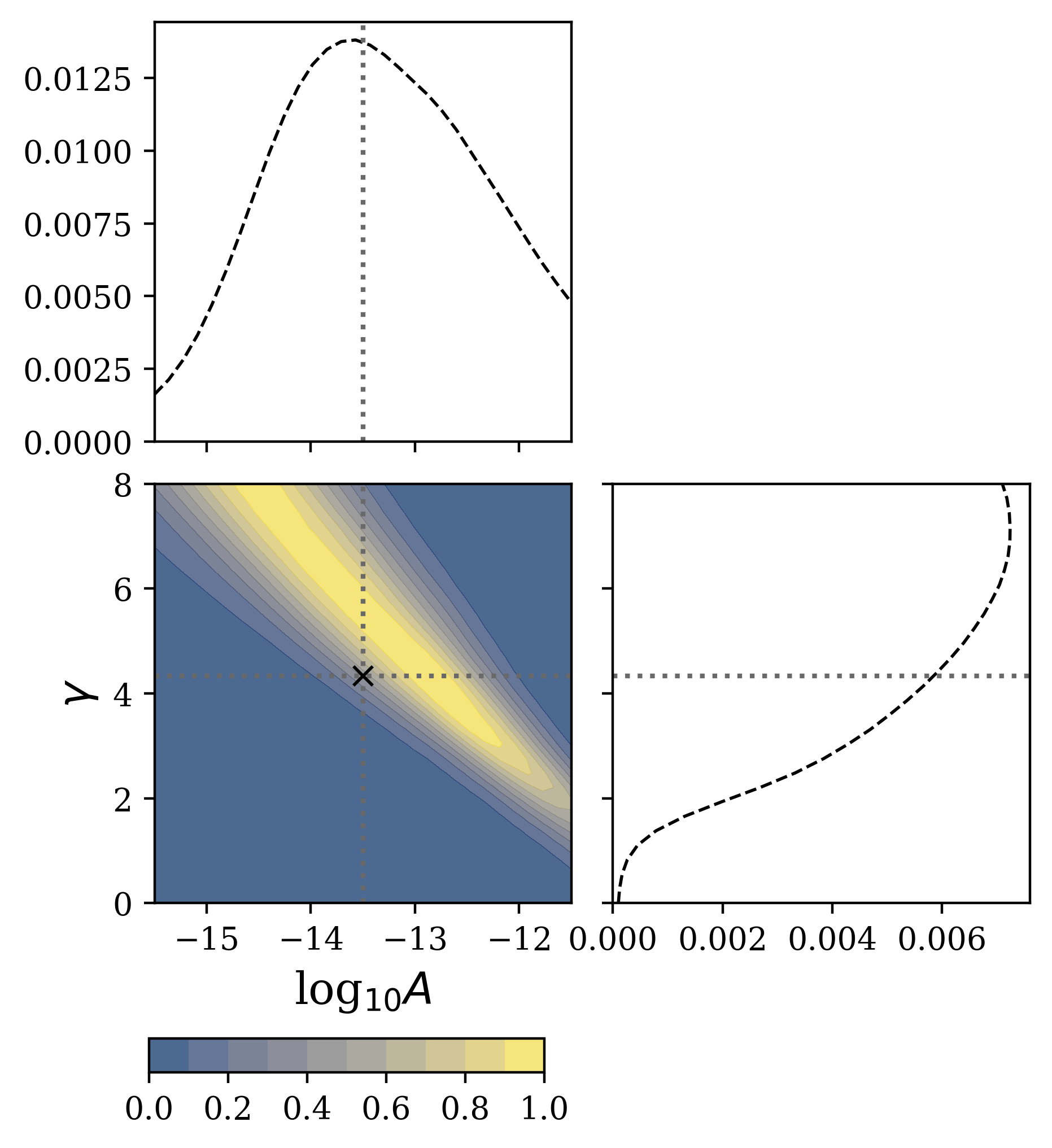}
    \caption{Posteriors of the GWB parameters averaged across 100 phase draws, evaluated using the trained conditional neural circular  spline flow with factorised frequency dependence for 30 pulsars and 12 frequency bins. The cross and the dashed lines indicate the injected GWB signal, with $\log_{10}\!A_\gwb^\mathrm{inj}=-13.5$ and $\gamma_\gwb^\mathrm{inj} = 13/3$.}
    \label{fig:avg_likelihood_30psrFRES}
\end{figure}

In addition, we notice that different realisations of the randomly drawn complex residuals lead to substantial variations of the likelihood shape and position. In the extreme, certain realisations lead to non-detections, or posteriors at the edge of the parameter plane. In order to evaluate this behaviour, we have averaged the posteriors of 100 phase samples. The result, shown in Fig.~\ref{fig:avg_likelihood_30psrFRES} indicates that despite the variations, on average, the phase likelihood evaluation is capable of reliably constraining the presence of a GWB signal.

\section{Frequency-resolved normalising flow}
\label{app:freq_res_flow}

A more realistic PTA data set requires jointly modelling the phase PDF across all frequency bins. For the 30-pulsar-8-bin PTA test case, this requires training a 240-dimensional NCSF conditioned on $(\log_{10}A_\gwb, \gamma_\gwb)$. The dimensionality of this problem, the large fraction of the parameter space across which the phases are uncorrelated, as well as the dynamic range of the correlation S/N make it challenging to find a flow setup that is capable of modelling the PDF, while training over reasonable time scales on a GPU. 

\begin{figure}
    \centering
    \includegraphics[width=\linewidth]{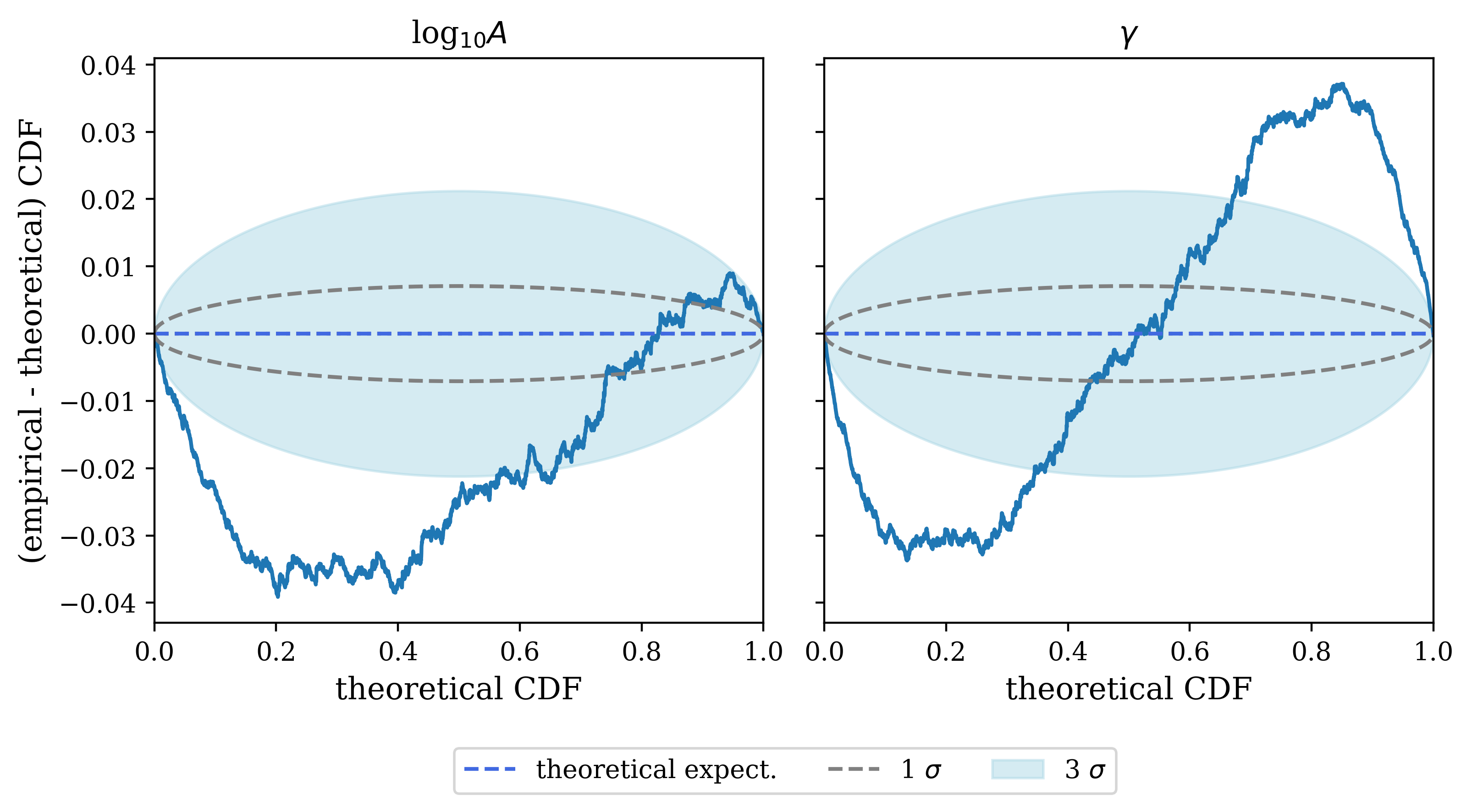}
    \caption{A $P-P$ plot of the recovered log-amplitude (left) and spectral index (right) recovered from 1000 evaluations of a trained conditional neural circular spline flow that includes all frequency bins simultaneously. }
    \label{fig:pp-plot_NSCF_p30b8}
\end{figure}

Our most successful architecture is a NCSF with 12 layers, in which each spline is fit using a $(32,32)$-featured perceptron. However, the p-p-plot shown in Fig.~\ref{fig:pp-plot_NSCF_p30b8} indicates that the flow is not properly representing the phase distribution. The deviations in the eCDF of the amplitude indicate that the mean of the flow distribution is shifted with respect to the mean of the actual phase PDF. The deviations in the eCDF of the spectral index indicate that the distribution is narrower than the actual PDF \citep{Lamb_2023}.

\begin{figure}
    \centering
    \includegraphics[width=\linewidth]{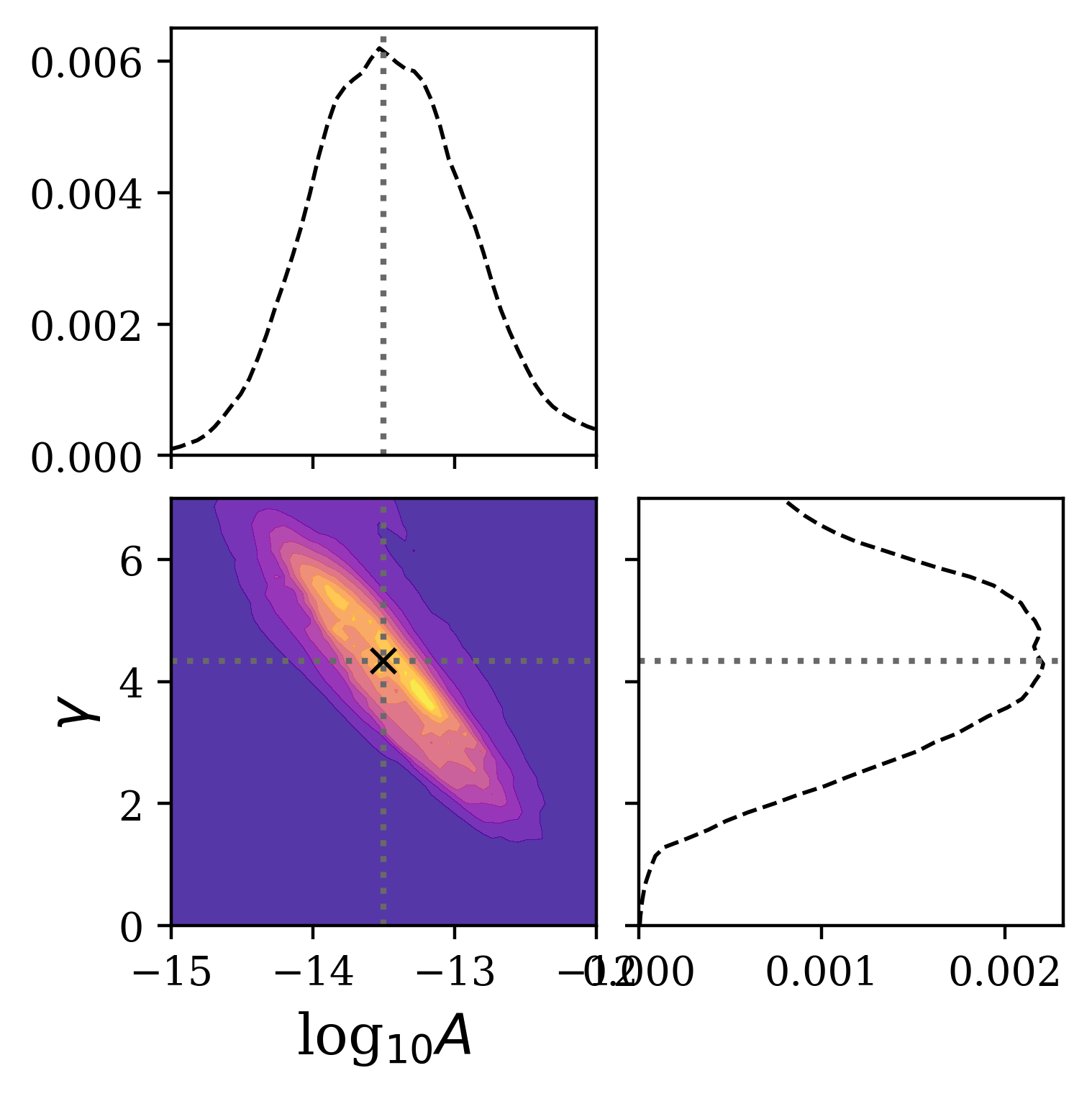}
    \caption{Exemplary evaluation of the empirical likelihood for using a randomly drawn set of phases using the trained conditional neural circular spline flow with simultaneous modelling of all frequency bins. The neural circular spline flow has a dimension of 240, corresponding to 30 pulsars and 8 frequency bins. The cross and the dashed lines indicate the injected GWB signal, with $\log_{10}\!A_\gwb^\mathrm{inj}=-14.0$ and $\gamma_\gwb^\mathrm{inj} = 13/3$.}
    \label{fig:llh_A_gamma_30psr8bin}
\end{figure}

\begin{figure}
    \centering
    \includegraphics[width=0.8\linewidth]{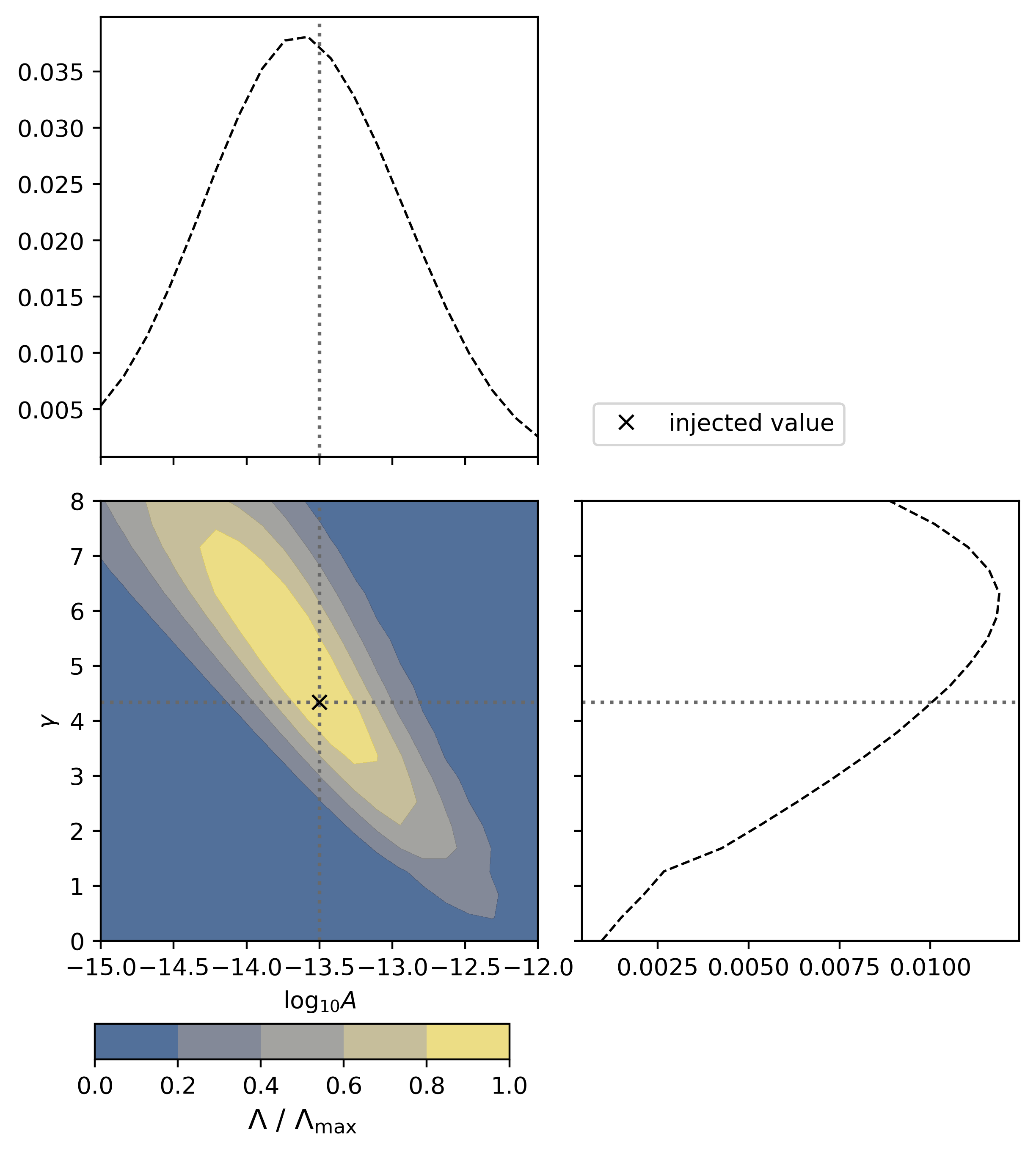}
    \caption{Posteriors of the GWB parameters averaged across 2000 phase draws, trained conditional neural circular spline flow with simultaneous modelling of all frequency bins. The flow has a dimension of 240, corresponding to 30 pulsars and 8 frequency bins. The cross and the dashed lines indicate the injected GWB signal, with $\log_{10}\!A_\gwb^\mathrm{inj}=-13.5$ and $\gamma_\gwb^\mathrm{inj} = 13/3$.}
    \label{fig:avg_likelihood_30psr8bin}
\end{figure}

Nonetheless, the flow model is sufficient to recover injected GWB parameters from individual phase draws. The LLH contours of the GWB spectral parameters for a single realisation of phases is shown in Fig.~\ref{fig:llh_A_gamma_30psr8bin}. Similarly to the frequency-factorised case, these posteriors strongly vary with different realisations of the randomly drawn complex residuals. Therefore, the averaged posteriors of 2000 phase samples are shown in Fig.~\ref{fig:avg_likelihood_30psr8bin}
Both results demonstrate that the higher-dimensional model is genuinely capable of constraining the spectral parameters of a GWB signal present in the phase correlations. Furthermore, the clearer decrease of the averaged distribution towards higher values of $\log_{10}A_\gwb$ and lower values of $\gamma_\gwb$ indicates that jointly evaluating the full phase distribution, i.e.\ explicitly modelling the frequency dependence, leads to better constrains on the GWB parameters. However, the averaged posterior reveals the impact of the mismatch between the true and modelled PDFs - the peak of the posterior is shifted towards a lower amplitude, yet steeper spectral index. Towards the analysis of realistic PTA data sets, these mis-modellings have to be overcome.

\end{document}